\documentclass[conference,a4paper]{IEEEtran}

\usepackage{amsmath}

\newcommand{\req}{\text{req}}
\newcommand{\mb}{\mathbf}
\newcommand{\mc}{\mathcal}
\newcommand{\F}{\mathcal{F}_{\text{req},t}}
\newcommand{\B}{\mathcal{B}}

\newcommand{\E}{\mathbf{E}}

\newcommand{\SINR}{\text{SINR}}

\usepackage{amsfonts,upgreek,pifont,amssymb,enumerate,color,algorithm,theorem}

\usepackage{algpseudocode}
\usepackage{cite,graphicx,subfigure,epstopdf}

\usepackage{mathtools}
\def\BibTeX{{\rm B\kern-.05em{\sc i\kern-.025em b}\kern-.08em
    T\kern-.1667em\lower.7ex\hbox{E}\kern-.125emX}}
\begin{document}
%
\title{Joint Long-Term Cache Allocation and Short-Term Content Delivery in Green Cloud Small Cell Networks}

\author{ 
Xiongwei~Wu$^1$, Qiang Li$^2$, Xiuhua Li$^{3,5}$, Victor C. M. Leung$^{4,5}$, P. C. Ching$^1$\\
$^1$Dept. of Electronic Engineering, The Chinese University of Hong Kong, Shatin, Hong Kong SAR, China\\
$^2$School of Info. \& Comm. Eng., University of Electronic Science and Technology of China, Chengdu, China\\
$^3$School of Big Data \& Software Engineering, Chongqing University, Chongqing, China\\
$^4$College of Computer Science \& Software Engineering, Shenzhen University, Shenzhen, China\\
$^5$Dept. of Electrical \& Computer Engineering, The University of British Columbia, Vancouver, Canada\\
E-mail: \{xwwu, pcching\}@ee.cuhk.edu.hk; lq@uestc.edu.cn; lixiuhua1988@gmail.com; vleung@ieee.org
}%
\maketitle 

\begin{abstract}
Recent years have witnessed an exponential growth of mobile data traffic, which may lead to a serious traffic burn on the wireless networks and considerable power consumption. Network densification and edge caching are effective approaches to addressing these challenges.
In this study, we investigate joint long-term cache allocation and short-term content delivery in cloud small cell networks (C-SCNs), 
where multiple small-cell BSs (SBSs) are connected to the central processor via fronthaul and can store popular contents so as to reduce the duplicated transmissions in networks. 
Accordingly, a long-term power minimization problem is formulated by jointly optimizing multicast beamforming, BS clustering, and cache allocation under quality of service (QoS) and  storage constraints. The resultant mixed timescale design problem is an anticausal problem because the optimal cache allocation depends on the future file requests. To handle it, a two-stage optimization scheme is proposed by utilizing historical knowledge of users' requests and channel state information. Specifically, the online content delivery design is tackled with a penalty-based approach, and the periodic cache updating is optimized with a distributed alternating method.
{Simulation results indicate that the proposed scheme significantly outperforms conventional schemes and performs extremely close to a genie-aided lower bound in the low caching region.}
\end{abstract}

\section{Introduction}



Rapid growth of mobile data traffic in recent years {has} imposed a serious traffic burden on the backhual and fronthaul of wireless networks. {Such a burden may generate considerable power consumption, which is one of major concerns in green wireless networks.} 
{To address these challenges}, deploying dense small cells and introducing edge caching are regarded as effective approaches towards fifth-generation wireless networks \cite{li2018resource}. 

By connecting multiple small-cell base stations (SBSs) to a central processor,  cloud small cell networks (C-SCNs) enable centralized optimization for signal processing and cooperation among a cluster of SBSs \cite{li2018resource}. {These benefits allow C-SCNs to achieve higher spectral efficiency and energy efficiency, compared with conventional cellular networks. For cache-enabled C-SCNs,} 
each SBS can prefetch popular contents during off-peak sessions. Generally, the requested contents from users are intensively correlated and dominated by several popular files \cite{golrezaei2012femtocaching}. The same content may be requested by users multiple times.
When users make requests, these cached contents are transmitted by local SBSs without consuming fronthaul resource. {Therefore, wireless edge caching can effectively reduce duplicated content downloads and thus offload network traffic.} This process substantially reduces power consumption. 

In such a cache-enabled C-SCNs, a fundamental problem is {\bf\emph{ how to design long-term caching strategies}} because content popularity may remain unchanged for a long time (in the order of hours or even days). Conventional caching schemes mainly focus on the cloud-to-BSs layer, and researchers have investigated offline strategies for cache allocation by assuming a priori distribution of content popularity, such as the Zipf distribution \cite{liao2017coding}. 
The authors in \cite{golrezaei2012femtocaching} proposed the Femtocaching scheme to achieve low latency. The authors in \cite{liao2017coding}  investigated offline maximum distance separable (MDS) coded caching schemes to minimize the average backhaul load by considering multicast and cooperation. 
A hierarchical edge caching and scheduling scheme was proposed to offload network traffic and reduce system costs in \cite{li2018hierarchical}. However, physical-layer transmissions were not considered in \cite{liao2017coding,golrezaei2012femtocaching,li2018hierarchical}. Cache resources should be judiciously scheduled so that they can be frequently reused for content delivery in a long period.  
Another fundamental problem is {\bf \emph{how to operate content delivery in a much shorter time-scale given the cache resource}}, i.e., considering signal processing and resource allocation for wireless transmissions. 
In \cite{wu2018content}, a delivery scheme for joint fronthaul content assignment and beamforming was investigated to achieve low latency. In \cite{tao2016content,peng2017layered}, joint multicast beamforming and BS clustering schemes were examined to minimize the delivery power.
All of these studies \cite{tao2016content,peng2017layered,wu2018content} only investigated short-term content delivery schemes under simple caching policies, such as caching entire contents and heuristic strategies for cache allocation. To our best knowledge, studies on joint long-term cache allocation and short-term content delivery {have not been well explored}. 


In this paper, we present a joint scheme for cache allocation, multicast beamforming, and BS clustering in C-SCNs.  Specifically, a mixed timescale optimization problem is formulated to minimize long-term power consumption for fronthaul and edge links under physical-layer transmission and storage constraints. {Since the optimal cache allocation depends on the future file requests, the resultant mixed timescale design problem is an {\it anticausal} problem. To handle it, inspired by \cite{xiang2017cross}, we develop a two-stage optimization scheme, which first utilizes the historical file requests to help optimize the cache allocation, and then based on the optimized cache allocation, multicast beamforming and BS clustering are jointly optimized in an online manner. Specifically,} for online content delivery, a penalty-based algorithm is proposed by applying the variational reformulation of binary constraints and convex-concave procedure (CCCP) technique to tackle the binary variables and Quality of Service (QoS) constraint. 
{In the simulation, we evaluate the performance of the proposed scheme under different network settings. Compared with selected benchmarks, the proposed scheme can achieve better power efficiency.}

The remainder of the paper is organized as follows. The system model is described in Section II.  The problem formulation is presented in Section III. In Section IV, we discuss the algorithms for implementing the proposed two-stage control scheme. The simulation results are presented in Section V and Section VI concludes the paper.

\section{System Model}
\subsection{Network Model}
\begin{figure}[h]
  \centering
  \includegraphics[scale=0.45]{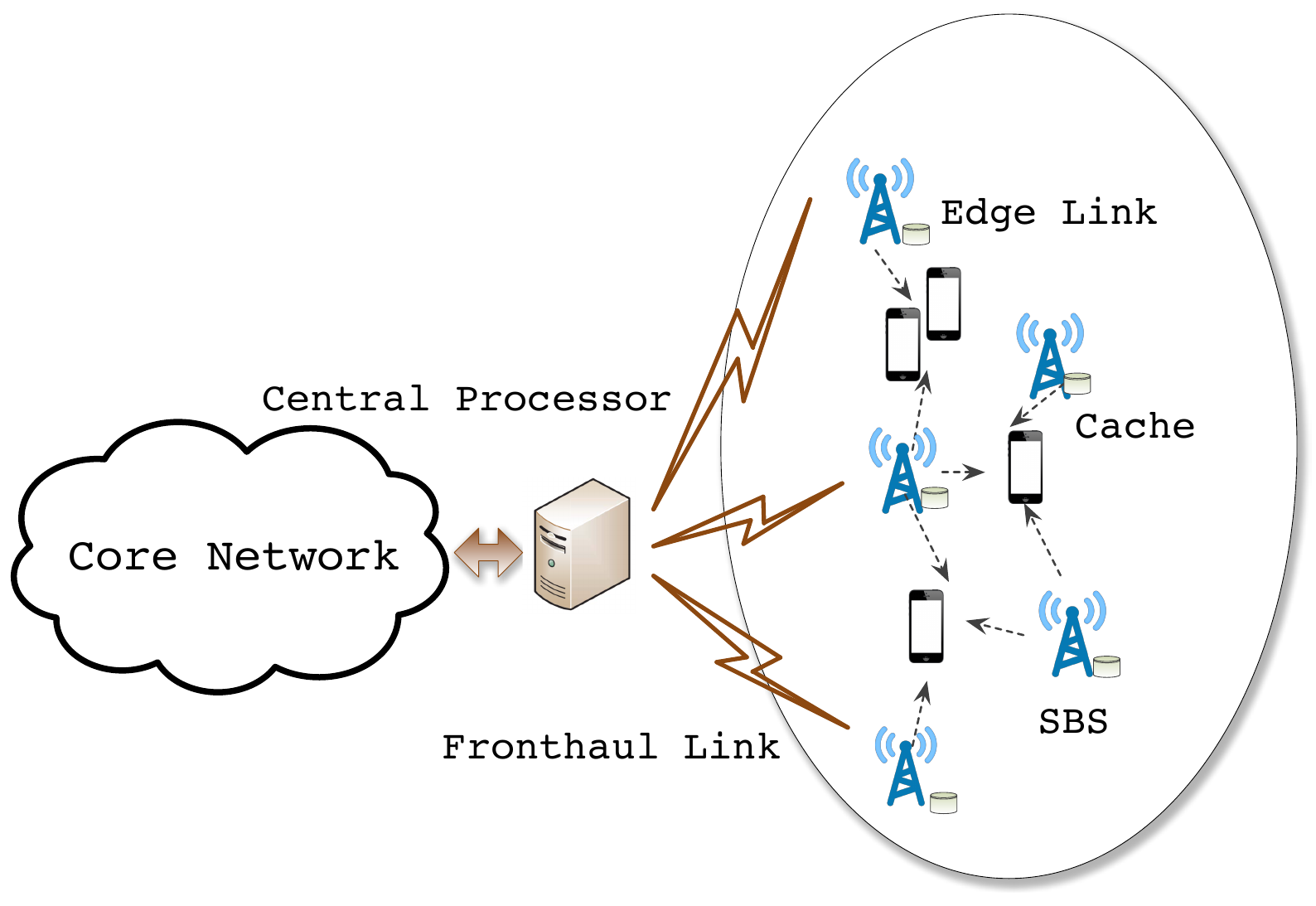}
  \caption{An illustration of cache-enabled C-SCNs. }
  \label{system}
\end{figure}

As illustrated in Fig. \ref{system}, we consider downlink cache-enabled C-SCNs, where $B$ densely deployed SBSs are connected to the central processor through wireless fronthaul links. Each SBS $b$ has $M$ antennas and a cache with the storage of $S_b$ bits. These SBSs cooperatively serve $K$ single-antenna users via wireless channels, which are referred to as the edge link. We assume that the cloud can access the entire content library, which stores $F$ files for delivery. 
The size of each file $f$ is $s_f$ bits. Let sets ${\cal{B}} = \{1,\dots, B\}$, $\mathcal{F} = \{1,\dots, F\}$, and $\mc K = \{1,\dots, K\}$ denote the indices of SBSs, files in the library, and users, respectively. In the proposed C-SCNs, the cached contents are reasonably scheduled for a long timescale $\mc T$, whereas the content delivery policy is designed in a much shorter term. Specifically, this system is time-slotted, and the duration of each time slot is assumed to be shorter than the wireless channel coherent time.

In each time slot, users may request files from the library through following certain content popularity distribution. For full generality, different users may have different preference patterns toward files \cite{liao2017coding}. Caching strategies for conventional caching designs were mainly investigated with  prior knowledge of preference patterns and content popularity distribution, such as the Zipf distribution \cite{liao2017coding}.
Specifically, users with the same preference pattern share a unique skewness parameter and file rank order. Accordingly, file $f$ is requested by users in preference pattern $i$ with probability  
\begin{align} 
    p_{i,f} = c_i \zeta_{i,f}^{-\kappa_i},  \label{distri} 
\end{align}
where $\zeta_{i,f}$ denotes the rank order of file $f$ in preference pattern $i$, $\kappa_i \geq 0$ is a skewness parameter, and $c_i$ is a normalization constant. However, in practice, these parameters in \eqref{distri} may not be reasonably justified as a priori knowledge, which makes it difficult for practical implementation. 

\subsection{Joint Cache Allocation and Content Delivery}
To allow the cached contents to be frequently reused for file transmission, we jointly consider long-term cache allocation and short-term content delivery. 
Let $\mb L= [l_{f,b}]$ denote the cache allocation matrix, where $0 \leq l_{f,b} \leq 1 $ denotes the fraction of file $f$ cached in BS $b$, $\forall f \in \mc F, b \in \mc B$. When users make requests from the cloud, file $f$ can be directly transmitted from the local SBSs if it is entirely cached. If the requested file is not fully cached in the local SBSs, the cloud is required to deliver the missing data fragments to the SBSs via fronthaul so that each SBS can recover the original file.
{Let $\mc F_{\req,t} \subseteq \mc F$ be the collection of requested files in the $t$th time slot. The users' requests in time slot $t$ are defined $\pi_t = \{(k,f)|k \in \mc K, f \in \F\}$. The users with the same requested file are grouped together and served by a cluster of SBSs in a multicast beamforming manner.}
Specifically, the users
in group $f$ are denoted as $\mathcal{G}_{f,t}$, and they are only requesting file $f$ in time slot $t$. 
With regard to cooperative delivery, a BS clustering matrix is defined as 
\begin{align} 
   \E_t = [ e_{f,b,t}] \in \{ 0,1 \}^{B \times F_{\req,t}}, \label{p1e}
\end{align} 
where ${F}_{\req,t}$ is the cardinality of $\mc F_{\req,t}$; the element $e_{f,b,t} = 1 $ indicates that SBS $b$ is selected to serve the multicast group $f$ in time slot $t$; otherwise, it is 0. 

For the edge link, 
the signal transmitted from SBS $b$ in time slot $t$ is 
    \begin{align}
      {{\mathbf{x}}_{b,t}} = \sum_{f' \in \F} {{{\mathbf{v}}_{f',b,t}}{x_{f',t}}}, 
    \end{align}
where $\mathbf{v}_{f,b,t} \in \mathbb{C}^M$ denotes the transmit beamformer from SBS $b$ for delivering file $f$, and the signal $x_{f,t} \in \mathbb{C}$ independently encodes file $f$ with distribution $ x_{f,t} \sim \mathcal{CN}(0,1)$, for any $t$. 
The transmit power of SBS $b$ is limited by $P_b$, i.e.,
\begin{align} 
  \sum_{f \in \F} \mb \|\mb v_{f,b,t}\|_2^2 \leq P_b, \forall b \in \B. \label{p1d} 
\end{align}  
Notably, if SBS $b$ does not transmit file $f$ in time slot $t$, the corresponding transmit beamformer $\mathbf{v}_{f,b,t}$ should be $\mathbf{0}$, i.e.,
\begin{align}
(1 - e_{f,b,t})\mathbf{v}_{f,b,t} = \mathbf{0}, ~\label{p1c}  \forall f\in \F, b\in \mc B
\end{align} %
Assuming a frequency-flat fading channel model, the received signal at  user {$k\in \mc G_{f,t} $} in time slot $t$ is given by
  \begin{align} 
     y_{k,t} = \underbrace{\mb h_{k,t}^H \mb v_{f,t} x_{f,t}}_{\text{desired signal}} + \underbrace{\sum_{f' \in \mc F _{\req,t} \backslash \{f\}} \mb h_{k,t}^H \mb v_{f',t} x_{f',t}}_{\text{inter-group~interference}} + z_{k,t}, 
  \end{align}   
where the integrated channel matrix ${{\mathbf{h}}_{k,t}} = \left[ {{\mathbf{h}}_{k,1,t}^H,{\mathbf{h}}_{k,2,t}^H, \cdots,{\mathbf{h}}_{k,B,t}^H}\right]^H$ and ${{{\mathbf{h}}_{k,b,t}}} \in \mathbb{C}^{M}$ denotes the channel matrix between BS $b$ and user $k$ in time slot $t$; the integrated beamformer ${{\mathbf{v}}_{f,t}} = \left[ {{\mathbf{v}}_{f,1,t}^H,{\mathbf{v}}_{f,2,t}^H,\cdots,{\mathbf{v}}_{f,B,t}^H} \right]^H$ denotes the beamformers from all SBSs to precode signal $x_{f,t}$; and ${z}_{k,t}$ denotes the additive complex Gaussian noise with distribution ${z_{k,t}} \sim \mathcal{CN}({0},\sigma_{k}^2)$. For notational convenience, we define set $\mc V_{t} = \{\mb v_{f,t}, f \in \F \}$.

Accordingly, the received signal-to-inference-plus-noise ratio (SINR) for user $k$ in time slot $t$ is given by
\begin{align} 
   \SINR_{k,t} = {|\mb h_{k,t}^H \mb v_{f, t}|^2}/g_{k,t}({\mc V_t}), 
\end{align}
where function 
$$g_{k,t}({\mc V_t}) \triangleq \sum_{f' \in \F \backslash \{f\}} |\mb h_{k,t}^H\mb v_{f',t} |^2 + \sigma_{k}^2.$$ 
Furthermore, the minimum received SINR for user $k$ in group $\mc G_{f,t}$ to successfully decode file is defined as $\gamma_{f}$, i.e., 
\begin{align} 
   \text{SINR}_{k,t} \geq \gamma_{f}, \forall k \in \mc G_{f,t}, \forall f \in \F. \label{p1b}
\end{align}  
Similar to \cite{tao2016content}, we consider a fixed and feasible transmission rate $R_f$ for file $f$, i.e., $R_f = B\log_2 (1 + \gamma_f)$, where $B$ (Hz) is the bandwidth of the edge link.  To avoid data interruption during content transmission, it requires minimum fronthaul capacity \cite{vu2018latency}
\begin{align} 
  R_{f,t}^{\text{FH}} = \sum_{b \in \B} ~ (1 - l_{f,b})e_{f,b,t}R_f,~ {f\in   \F} .
\end{align} 

\subsection{Delivery Power Consumption}


The power consumption mainly arises from content delivery through fronthaul and edge links. In particular, the edge transmission power is calculated by
    $P_{E,t} = \textstyle\sum_{f \in \F, b\in \B} \delta_b \|\mb v_{f,b,t}\|^2,$
  where $\delta_b$ is a slope of the load-dependent power constant for SBS $b$. 
And the fronthaul transmission power is calculated by
  $ 
    P_{F,t} = \textstyle\sum_{f \in \F} \sum_b \beta( 1 - l_{f,b})R_fe_{f,b,t}
  $, 
  where $\beta$ denotes the energy efficiency of the fronthaul link \cite{peng2017layered}. Therefore, the power consumption for time slot $t$ is given by 
  \begin{align}
    & P_t (\mb L, \mc V_t, \mb E_t) \notag\\
     =&   \sum_{f \in \F,b \in \mc B} \delta_b \|\mb v_{f,b,t}\|^2 + \sum_{f \in \F, b \in \mc B} \beta( 1 - l_{f,b})R_fe_{f,b,t}
  \end{align}

\section{Problem Formulation}
\subsection{Mixed Timescale Problem}
\begin{figure}[h]
	\centering
	\includegraphics[scale=0.40]{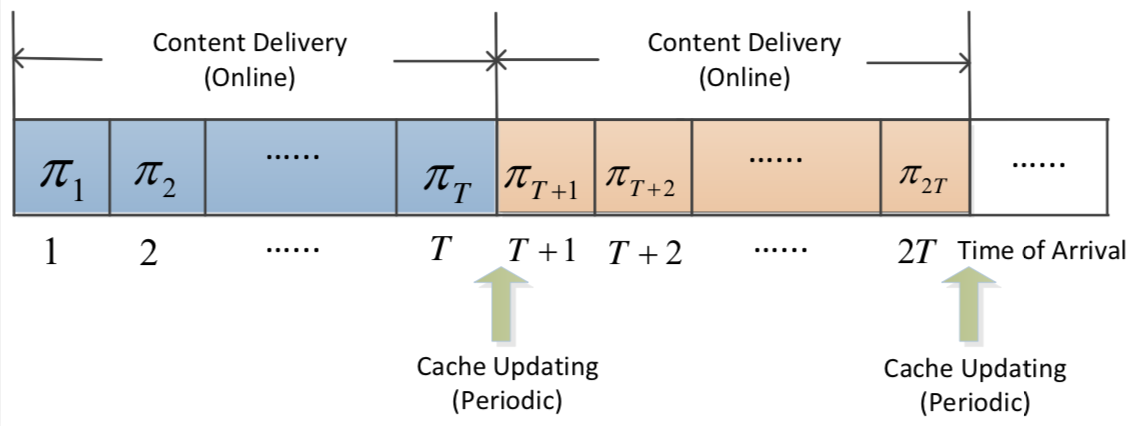}
	\caption{A mixed timescale model for periodic cache updating and online content delivery. }
	\label{JointScheme} 
\end{figure}

{Consider that all SBSs' caches are periodically updated every $T$ time slots; see Fig.~\ref{JointScheme}. Our goal is to design the cached files at each SBS, i.e., $\mb L$, so that the total content delivery  power consumption 
during the next $T$ time slots is minimized. Since the transmission power in the next $T$ time slots  also depends on the SBS clustering $\mc E = \{\mb E_t, t \in \mc T\}$ and the beamforming $\mc V = \{ \mc V_t, t\in \mc T\}$ with $\mc T = \{t_0+1, \cdots, t_0+T\}$ and $t_0$ being the current cache-updating time slot, it is necessary to jointly optimize the large timescale variable $\mb L$ and the small timescale variables $(\mc E,\mc V)$, which leads to the following mixed timescale power minimization problem: }
\begin{subequations} \label{eq:main_problem}
  \begin{align}
    \min_{{\mathbf{L}},{{\mathcal{V}}},{{\mathcal{E}}}}~& 
    \sum_{t \in \mc T}  {P_t(\mb L,{{\mathcal{V}}},{{\mathcal{E}}})}, \\
    s.t. ~&  \sum_{f \in \mc F} l_{f,b}s_f \leq S_b, ~ b \in \B, \label{b}\\
    & 0 \leq l_{f,b} \leq 1, ~\forall f \in \mc F, b \in \mc B, \label{c}\\
    &  {\eqref{p1e}, \eqref{p1d}, \eqref{p1c}, \eqref{p1b}}, \forall t~\in \mc T , \label{d}
\end{align}
\end{subequations}  
where {$s_f>0$ denotes the size of file $f$ (in bits); $S_b>0$ represents the maximum storage (in bits) of SBS $b$}; constraints \eqref{b} and \eqref{c} are used to avoid storage overflow. {Notably, problem \eqref{eq:main_problem} is {\it anticausal} at the beginning of the time slot $t_0$ because the users' requested files $\F, t\in \mc T$ and their CSI in the next $T$ time slots are generally unknown.} 
 
\subsection{Problem Decomposition for Two-Stage Updating Scheme}
{To circumvent the difficulty of anticausality of problem~\eqref{eq:main_problem}, we observe that in practice, the users' preference and the content popularity distribution should vary slow, and users' CSI is usually temporally correlated (say in slow fading scenario) \cite{xiang2017cross}. Therefore, by appropriately making use of the 
historical data up to $t_0$, we may get a good prediction of the future requests. Subsequently, we propose a two-stage design approach to handle problem~\eqref{eq:main_problem}.


\subsubsection{Cache updating design} This is implemented by the end of previous transmission block. At this stage, the historical users' requests and CSI can be gathered in the cloud. Therefore, cache allocation matrix $\mb L$ is obtained by solving the following problem $\mc P$. 
  \begin{align}
    \mc P: \min_{{\mathbf{L}},{{\mathcal{V}}},{{\mathcal{E}}}}~& 
    \sum_{t \in \mc T'} {P_t(\mb L,{{\mathcal{V}}},{{\mathcal{E}}})}, ~~ 
    {\rm s.t.} ~ \eqref{b} - \eqref{d}, 
\end{align}
{where we recall $\mc T' = \{t_0 - T +1, t_0 - T +2, \ldots, t_0 \}$ in $\mc P$.}
\subsubsection{Online content delivery design} The solution $\mb L^*$ to problem $\mc P$ is used for cache allocation in block $\mc T$. 
In each time slot $t \in \mc T$, the cloud coordinates how to deliver requested files based on the cached contents in SBSs $\mb L^*$, real-time users' requests $\pi_t$, and instantaneous CSI. Consequently, the transmission policy $(
\mc V_t, \mb E_t)$ is optimized by solving the following power minimization problem:
\begin{align} 
   \mc P_0:\min_{ \mc V_t, \mb E_t}~& {P_t(\mb L^*,{{\mathcal{V}}}_t,{{\mathbf{E}}}_t)} ~~~~\notag\\{\rm s.t.}~&{ \eqref{p1e}, \eqref{p1d}, \eqref{p1c}, \eqref{p1b}}
\end{align} 
for $t=t_0+1, \ldots, t_0 +T$. 

\section{Proposed Algorithms}
For ease of discussion, we first tackle the online content delivery problem $\mc P_0$. Obtaining an optimal solution is NP hard due to BS clustering and minimum SINR requirement. To make this problem more tractable, we introduce a variational reformulation of the binary constraint. 

\subsection{Variational Reformulation}
To address the discontinuity in $\mc P_0$, the binary constraint \eqref{p1e} can be recast into $e_{f,b,t}^2  -e_{f,b,t} = 0$, 
which can be equivalently rewritten as two continuous constraints
\begin{align} 
  e_{f,b,t}^2 - e_{f,b,t} \geq 0, \forall f \in \F, b \in \mc B, \label{h}\\
  e_{f,b,t}^2 - e_{f,b,t} \leq 0, \forall f \in \F, b \in \mc B. \label{i}   
\end{align}
The QoS constraint \eqref{p1b} can be transformed as
\begin{align} 
   \gamma_{f}  g_{k,t}({\mc V_t}) - |\mb h_{k,t}^H \mb v_{f,t}|^2 \leq 0, \forall k \in \mc G_{f,t}, \forall f \in \F, \label{l}
\end{align}
where the difference of two quadratic functions can be found on the left hand side.  Moreover, the equilibrium constraint \eqref{p1c} can be equivalently written as
\begin{align} 
  \|\mb v_{f,b,t}\|_2 \leq e_{f,b,t}\sqrt{P_b}, \forall f \in \F, b \in \mc B,  \label{m}
\end{align}
which is convex. When $e_{f,b,t} = 0$, the associated beamformer $\mb v_{f,b,t} = 0$; otherwise, $e_{f,b,t} = 1$, the feasible beamformer $\mb v_{f,b,t}$ is determined by constraint \eqref{p1d}.

By fixing cache status $\mb L = \mb L^*$, problem $\mc P_0$ can be equivalently reformulated as
        \begin{align}
            \notag
            \mc P_1 (\mc V_t, \mb E_t):\min_{ \mc V_t, \mb E_t}~&  {P_t(\mb L^*,{{\mathcal{V}}}_t,{{\mathbf{E}}}_t)}
            \\
           {\rm s.t.} ~& \eqref{p1d}, \eqref{h} - \eqref{m}, \notag
        \end{align}
which is a general difference of convex (DC) program because the pattern of the DC function can be found in constraints \eqref{h} and \eqref{l}. Hence, problem $\mc P_1$ can be efficiently solved through the CCCP technique \cite{lipp2016variations}. 

\subsection{Proposed Penalty CCCP Online Delivery}
Notably, the tight coupling between constraints \eqref{h} and \eqref{i} may make CCCP iterations infeasible. Therefore, we introduce a penalty CCCP online delivery algorithm to address this difficulty. 

Inspired by the idea in \cite{lipp2016variations}, we introduce other nonnegative slack variables $\Omega_t = \{\omega_{f,b,t}\}$ to relax constraint \eqref{h} as follows
\begin{align} 
  e_{f,b,t} - e_{f,b,t}^2 \leq \omega_{f,b,t}, \forall f \in \F, b \in \mc B. \label{t}
\end{align} 
The violation is further penalized by minimizing the sum of slack variables. Hence, problem $\mc P_1$ is relaxed as
\begin{subequations}
        \begin{align}
            \mc R_0:\min_{ \Theta_t}~&  {P_t(\mb L^*,{{\mathcal{V}}}_t,{{\mathbf{E}}}_t)} + \lambda\sum_{f,b} \omega_{f,b,t}, \label{r0a}\\
            s.t. ~& \omega_{f,b,t} \geq 0, \forall f \in \F, b \in \mc B,  \label{r0b}\\
            & \eqref{p1d}, \eqref{i} - \eqref{t},  \label{r0c}
        \end{align}
\end{subequations}
where variables $\Theta_t = \{\Omega_t, \mc V_t, \mb E_t\}$ and the penalty parameter $\lambda \geq 0$. Notably, an optimal solution $\mb E_t^*$ meets the condition of $\omega_{f,b,t} = 0, \forall f,b$. Hence, we can always gradually increase $\lambda$ so that the slack variable $\omega_{f,b,t}$ sufficiently approaches 0, resulting in a binary solution.  

On the basis of the principle of CCCP, constraints \eqref{l} and \eqref{t} are approximated by introducing their first-order Taylor expansions at the local point. We tackle problem $\mc R_0$ by solving the following inner approximation successively 
          \begin{align*}
           \mc R_0^{(i)}: &\min_{ \Theta_t}~~ {P_t(\mb L^*,{{\mathcal{V}}}_t,{{\mathbf{E}}}_t)} + \lambda\sum_{f,b} \omega_{f,b,t}\\
           {\rm s.t.}~~& (e_{f,b,t}^{(i)})^2  + (1 - 2 e_{f,b,t}^{(i )} )e_{f,b,t}  \leq \omega_{f,b,t}, \forall f, b,\\
            & \gamma_{f}g_{k,t}(\mc V_t) -  2\text{Re} \big\{ \mb v_{f,t}^{(i)H} \mb h_{k}\mb h_k^H \mb v_{f,t}\} \notag\\&~~+\big|{\mb h _k^H \mb v_{f,t}^{(i)}}\big|^2 \leq 0,\forall k \in \mc G_{f,t}, \forall f \in \F, \\
            &~~\eqref{p1d}, \eqref{i}, \eqref{m}, \eqref{r0b},
        \end{align*}
where the local points $\mb v_{f,t}^{(i)}$ and $e_{f,b,t}^{(i)}$ are the solution in the last iteration.  Notably, problem ${\mc R_0^{(i)}}$ is convex and can be efficiently solved via standard solvers, such as CVX \cite{grant2008cvx}. 

\begin{algorithm}[h]
\caption{Penalty CCCP Online Delivery Design  }
\begin{algorithmic}[1]\label{AL1}
\State {\bf Initialize} $i =0$ , $ \mc V_t^{(0)}, \mb E_t^{(0)}$, $\lambda > 0, \tau >1, \lambda_{\max}$
\State  {\bf Repeat}
\State ~~~~Solve convex problem $\mc R_0^{(i)}$ to obtain $\mc V_t^{(i+1)}, \mb E_t^{(i+1)}$
\State ~~~~Update $\lambda \leftarrow \min\{\lambda_{\max}, \tau\lambda\}$
\State ~~~~Update $i \leftarrow i+1$  
\State {\bf Until} {{some} stopping criterion is satisfied}
\end{algorithmic}
\end{algorithm}

The entire procedure is presented as Algorithm 1. To find a good initial point, we can always assign a small value for the initial penalty parameter. 
The parameter $\lambda_{\max}$ is used to prevent numerical issues. 
When the penalty parameter $\lambda = \lambda_{\max}$, the sequence of the objective value of problem $\mc R_0$, generated by Algorithm 1 in each iteration, converges based on the principle of CCCP \cite{lipp2016variations}. 

\subsection{Proposed Alternating Design for Cache Updating}
For cache updating problem $\mc P$, constraints \eqref{b} and \eqref{c} are linear. Therefore, problem $\mc P$ can also be directly solved by the CCCP technique. However, the dimension of variables and the number of constraints are in $\mc O(T)$ of that of problem $\mc P_0$. This fact makes numerical implementation difficult, due to the excessively high computation complexity. In the following subsection, problem $\mc P$ is decoupled into multiple subproblems, and a parallel algorithm is proposed accordingly.

When cache allocation matrix $\mb L$ is fixed, problem $\mc P$ can be decomposed into a group of independent subproblems $\mc P_1 (\mc V_t, \mb E_t), {\forall t \in \mc T'}$. Hence, we leverage the alternating  method to solve problem $\mc P$. Specifically, by fixing $\mb L = \mb L^{(i)}$, the first block $\{\mc V, \mc E\}$ can be updated by solving a group of subproblems $\mc P_1 (\mc V_t, \mb E_t), {\forall t \in \mc T'} $ in parallel. For the other block $\mb L$, by fixing $\{\mc V, \mc E\} = \{ \mc V^{(i)}, \mc E^{(i)}\}$, it can be updated by solving the following linear program
\begin{align} 
 \mc P_2: \min_{\mb L}~ \sum_{t \in \mc T'}P_t (\mb L, \mc V_t^{(i)}, \mb E_t^{(i)}) ~~{\rm s.t.}~~ \eqref{b}, \eqref{c}. 
\end{align}  
The entire procedure is shown as Algorithm 2.
\begin{algorithm}[h]
\caption{Periodic Cache Updating Design}
\begin{algorithmic}[1]\label{AL1}
\State {\bf Initialize} $i =0$ , $ \mb L^{(0)}$
\Repeat
\State Fixing $\mb L = \mb L^{(i)}$, solve problems $\mc P_1(\mc V_t, \mb E_t)$ for ${\forall t \in \mc T'}$ in parallel by Algorithm 1, and the solution is defined as $\{\mc V^{(i+1)}, \mc E^{(i+1)}\}$. 
\State Fixing $\{\mc V, \mc E\} = \{\mc V^{(i+1)}, \mc E^{(i+1)}\}$, solve problem $\mc P_2$ and the solution is defined as $\mb L^{(i + 1)}$
\State  Update $i \leftarrow i+1$  
\Until {some stopping criterion is satisfied}
\end{algorithmic}
\end{algorithm}

\section{Performance Evaluation}
In the section, we present the simulation results to evaluate the performance of the proposed two-stage control in practical scenarios. Consider cache-enabled C-SCNs, in which a total of $7$ densely deployed SBSs are randomly distributed within the macro cell. The macro cell covers a hexagonal area with an edge length of 1 km.   
Each SBS has four antennas and the same fractional caching capacity $\mu$, where $\mu = \sum_{b \in \B}S_b/(B\sum_{f \in \mc F} s_f)$. Multiple single-antenna users are randomly distributed within the macro cell, except for the circle of 30 m around each SBS. For each time slot, the wireless channels are modeled as large-scale fading and small-scale fading. Specifically, as in \cite{tao2016content}, the path-loss $PL(\text{dB}) = 148.1 + 37.6\lg(d)$, where $d$ is the distance between SBSs and users in kilometers; the antenna power gain is 10 dBi and the log-normal shadowing parameter is $8$ dB; and the small scale fading is Rayleigh fading with unit variance. For ease of discussion, we assume that  active users are quasi-static. Thus, large-scale fading does not change over the transmission blocks, but small scale fading is random and independent in each time slot. 
The default scenario is as follows: 100 popular files are in the library, the fractional caching capacity is 0.2, the maximum transmit power for each SBS is 40 dBm, the minimum received SINR for each file is $5$ dB,  each transmission block contains 100 time slots (simulation trials). Without loss of generality, the file preferences of the active users have $I = 4$ patterns. For each preference pattern, the skewness parameter is uniformly selected from a continuous set [1,2], and the associated file rank order is randomly generated \cite{liao2017coding}. Four active users are considered for each preference pattern (if not specified otherwise). For the parameters of the power consumption model, we set the typical values $\delta_b = 4$ and $\beta = 10^{-7}$\cite{peng2017layered}.

\begin{figure}[h]
  \centering
  \includegraphics[scale=0.38]{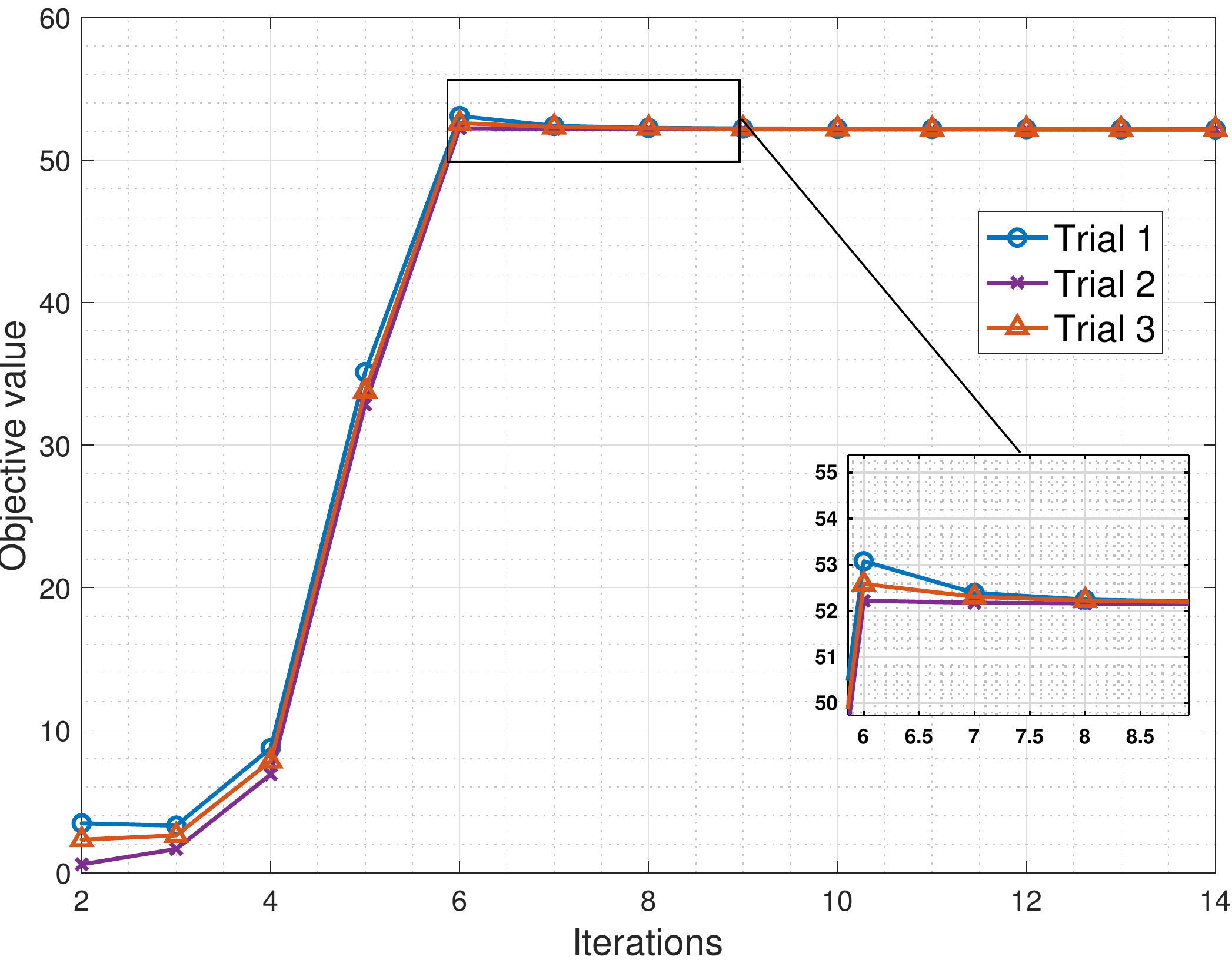}
  \caption{Convergence behavior of the proposed penalty CCCP online delivery design.}
  \label{AC}
\end{figure}
We show the convergence behavior of the proposed penalty CCCP online delivery design in Fig. \ref{AC}. 
Three independent trials are considered. The initial points of beamformers are randomly generated in each trial. For comparison, we set the initial penalty value $\lambda = 1$ and $\lambda_{\max} = 1000$, the increasing factor is $\tau = 5$, and $e_{f,b,t} = 0.1, \forall f,b$.  Cache allocation $\mb L$ is given by the proposed caching scheme.
The objective value is first lifted within six iterations due to the increase in the penalty value in the objective function. After $\lambda$ reaches the maximum value, the objective values decrease gradually. All three trials almost converge to the same value. These results indicate that the proposed algorithm for online beamforming can converge within 10 iterations and is not sensitive to the initial value for beamforming vectors.  

In the following, we compare the performance of the proposed caching policy with the following benchmarks:
\begin{itemize}
  \item {\bf Uniform Caching (UC):} Each SBS uniformly and randomly fetches a fraction $\mu$ of each file in the library  \cite{park2016jointJ}, i.e., $l_{f,b} = \mu, \forall f,b.$ This scheme serves as an upper bound for performance evaluation.
 
  \item {\bf Genie-Aided Caching (GAC):} Assuming that we have already obtained the users' requests and CSI in the future block, the cache updating is scheduled by the proposed approach. This scheme serves as a lower bound for performance evaluation.
  \item {\bf Least Recently Used (LRU):} Similar to \cite{cao1997cost}, this online scheme is implemented according to the recently accessed frequency of contents. When the current request is not available in local SBSs, we evict the content that was accessed least and cache the newly arrived content. 
\end{itemize}
\begin{figure}[h]
  \centering
  \includegraphics[scale=0.33]{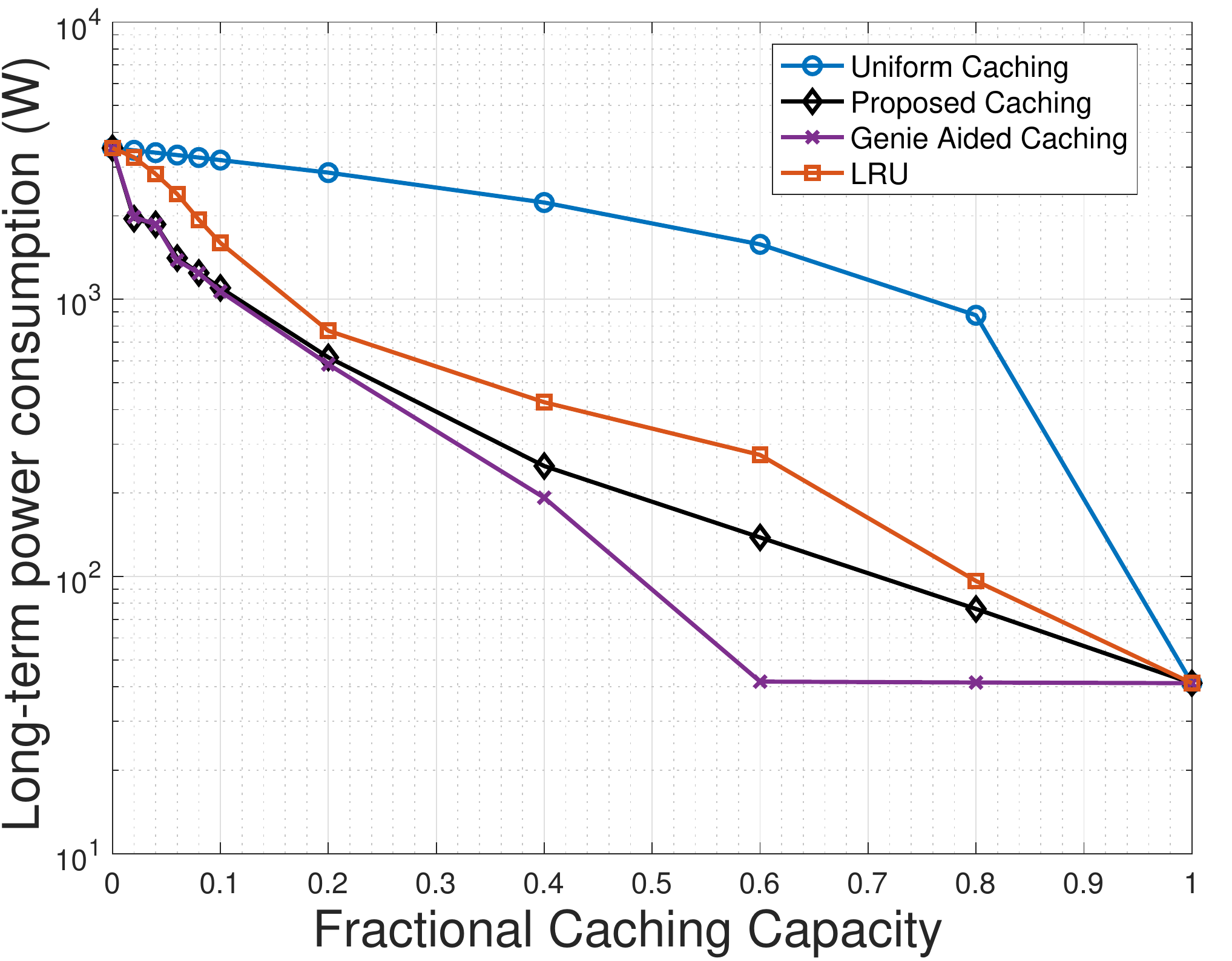}
  \caption{Long-term power consumption versus fractional caching capacity.}
  \label{FHC}
\end{figure}

We illustrate the impact on fractional caching capacity on long-term power consumption in Fig. \ref{FHC}. We observe that in the low caching region ($\mu\leq 0.2$), when the cache capacity increases, the averaged total power consumption degrades dramatically. This finding indicates that the total power consumption is dominated by content delivery in the fronthaul link. In particular, the proposed caching policy obtains results that are extremely close to those of GAC in the low caching region. Moreover, the proposed design consumes $91.2 \%$ and $49.6 \%$ less power than UC and LRU over the entire horizontal axis, respectively.
\begin{figure}[h]
  \centering
  \includegraphics[scale=0.38]{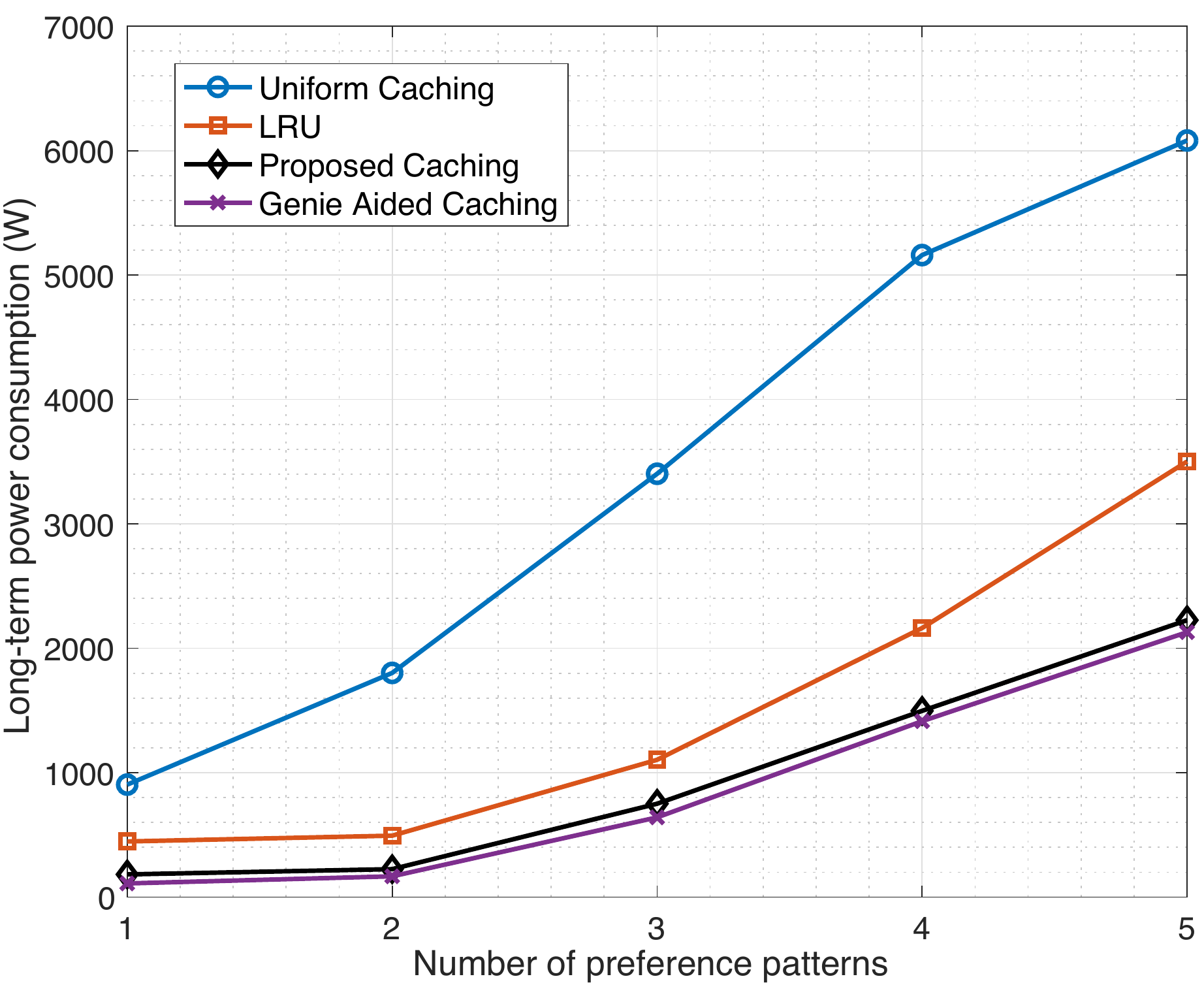}
  \caption{Impact of the number of users' preferences patterns.}
  \label{KK}
\end{figure}
In Fig. \ref{KK}, we investigate the performance of the proposed scheme by varying the number of users' preference patterns. The proposed scheme achieves power consumption that is very close to GAC. As the number of preference patterns increases, the gaps between the curves of the proposed method and conventional ones become larger. The reason is that the cached contents of the proposed scheme are adaptive to  users' content popularity and  clustering patterns of SBSs by utilizing historical users' requests and CSI. Intuitively, the cached contents of SBSs can always be locally hit by users with high probability. Frequent reuses of the cached contents in local SBSs lead to a reduction in power consumption.  

\section{Conclusion}
We have developed a two-stage optimization scheme for joint cache allocation, multicast beamforming, and BS clustering to save power consumption in C-SCNs without the priori distribution of the content popularity. 
For practical implementation, the content delivery design was tackled by a penalty-based algorithm by using a variational reformulation of the binary constraints. An alternating algorithm with parallel implementation has also been proposed for the periodic cache updating. The simulation results have revealed the superior performance of the proposed scheme. 
\bibliographystyle{IEEEtran}
\bibliography{references}
\end{document}